\documentstyle[prl,aps,epsf,psfig]{revtex}
\begin{document}
\twocolumn[\hsize\textwidth\columnwidth\hsize
           \csname @twocolumnfalse\endcsname
\title{Electric field in stationary superconductors}
%      ============================================
\author{Jan Kol\'a\v cek, Pavel Lipavsk\'y}
\address{Institute of Physics, ASCR, Cukrovarnick\'a 10,
16253 Prague 6, Czech Republic}
\maketitle
\begin{abstract}
%================
It is generally accepted that vortex core is charged, what illustrates
that even in stationary superconductors electric field may be present. 
Vortex structure and other properties of superconductors are usually 
calculated in the framework of the Ginzburg-Landau theory, which does 
not cover electric field. We show that the generalization of the 
GL theory due to Bardeen allows one to derive a third GL equation for 
the electrostatic potential. Since the Bardeen theory applies to all 
temperatures, the presented theory enables one to calculate charge 
profiles of the vortex under quite general conditions. The theory is
consistent with the BCS-Gorkov results.
%\vspace{1pc}
\end{abstract}
\maketitle
    \vskip2pc]

\section{Introduction}
%======================
As was pointed out by Adkins and Waldram \cite{AW68} and later 
followed by others, see e.g. \cite{Hong}, a difference of the 
chemical potential in a superconducting versus normal state, 
is compensated by the electrostatic potential. If the gap is
modulated in space, say due to supercurrents, this potential
results in an internal electric field. This idea has been used
later to estimate the charge of vortices \cite{Blatter96,95Khomskii}. 
For superconductor of the first kind, the presence of nonzero 
electric field in the superconductor carrying tranport current 
was first experimentally proved by Bok and Klein \cite{68Bok}
who measured the surface charge. For recent measurement with 
this method see e.g. \cite{96Chiang}. A new experiment, which
allows one to measure the electric field in the bulk of the
superconductor of the second kind in the mixed state, has been
performed by Kumagai, Nozakii and Matsuda \cite{00Matsuda}.

A vortex structure is usually calculated in the framework of 
the Ginzburg-Landau (GL) theory, which is applicable only 
for temperatures close to the critical temperature $T_c$ and 
in its standard formulation does not cover the electric field. 
The electric field can be evaluated, however, from the Poisson
equation, called by Jakeman and Pike \cite{JP67} the third GL
equation, to reflect that the source term (density of unscreened
charge) is a functional of the GL function.

Various expressions for the source term of the third GL 
equation can be found in the literature
\cite{AW68,Hong,Blatter96,95Khomskii,JP67,Rickayzen,KLB01}, 
most of contributions are of similar amplitude and seem to 
be merely alternative approximations of the same mechanism.
This is not, however, the case. There are at least three
distinguishable mechanisms which create the electric field 
in superconductors. To clarify this point, we derive the
third GL equation with all these mechanisms from a simple 
phenomenologic theory of the GL type. Not to be restricted 
to temperatures close to $T_c$, we adopt Bardeen's 
extension of the GL theory \cite{Bardeen1} based on the
Gorter-Casimir model \cite{Gorter1}. 

\section{Gorter-Casimir two fluid model}
%======================
The free energy of the normal state metal without electric
and magnetic fields reads
\begin{eqnarray}
{\mathcal F}_{\rm n} = U - {1\over 2} \gamma T^2,
\end{eqnarray}
where $U$ is internal energy including lattice vibrations 
and $\gamma T$ is the electronic specific heat.
According to the Gorter-Casimir two fluid model, the free energy
of a superconducting state without fields can be written as
\begin{eqnarray}
{\mathcal F}_{\rm s} = U - {1\over 4} \gamma T_c^2 \omega 
      -{1\over 2} \gamma T^2 \sqrt{1-\omega}.
\end{eqnarray}

Equilibrium value of the order parameter 
$\omega$ makes $\mathcal F_{\rm s}$ minimum. From the condition
${\partial {\mathcal F}_{\rm s} \over \partial \omega}=0$
it follows that   
\begin{eqnarray}
\omega_{\rm eq} = 1 - t^4 ,
\end{eqnarray}
where $t=T/T_{\rm c}$ is the reduced temperature.
At $T=0$ the equilibrium value $\omega_{\rm eq}=1$ and 
${\mathcal F}_{\rm n}-{\mathcal F}_{\rm s}
     = {1\over 4} \gamma T_{\rm c}^2 
     = {1\over 2} \mu H_{\rm c}^2 $, where $H_{\rm c}$
is the thermodynamic critical field.
At $T=T_{\rm c}$ the equilibrium value of $\omega$ is zero and 
${\mathcal F}_{\rm s}={\mathcal F}_{\rm n}
     = U-{1\over 2} \gamma T_{\rm c}^2 $
as it should. According to its temperature dependency, the
order parameter $\omega$ can be identified with the square
of the GL effective wave function $\psi$, normalized to 
the superfluid fraction, $|\psi|^2 ={n_{\rm s}\over n}$,
where $n_s$ is the superconducting charge carriers density and
$n$ is the total density of charge carriers. 

Near the critical temperature the order parameter $\omega$
is small. Using the expansion 
$\sqrt{1-\omega}\approx 1-{1\over2}\omega-{1\over8}\omega^2$
the Gizburg-Landau free energy,
$\alpha\left|\psi\right|^2+{1\over2}\beta\left|\psi\right|^4$
is recovered, where the parameters $\alpha$ and $\beta$ 
in the used normalization are
$\alpha=- 2\mu H_{\rm c}^2 (1-t)/n$ and
$\beta=\mu H_{\rm c}^2/n^2$.  Accordingly, one can see the
free energy of Ginzburg and Landau as an asymptotic form of
the more general Gorter-Casimir model.

\section{Bardeen's extension of the GL theory}
%======================
To extend the region of applicability of the Ginzburg-Landau theory,
Bardeen \cite{Bardeen1} replaced the Ginzburg-Landau polynomial free
energy by the free energy due to Gorter and Casimir.
The free energy then reads
\begin{eqnarray}  \label{FreeEnergy}
%                 ==================
{\mathcal F}&=& U
  -{1\over 4}\gamma T_c^2 \left|\psi\right|^2 
  -{1\over 2}\gamma T^2 \sqrt{1-|\psi|^2}
  +{{\bf B}^2\over2\mu}
\nonumber \\
  &+&{n\over 2}{\left|(i\hbar\nabla+e^*{\bf A})\psi\right|^2\over 2m^*}
  -\epsilon {{\bf E}^2\over 2}
  +\varphi\rho.
\end{eqnarray}
We note that the last two terms, which describe the electrostatic 
field, have not been assumed by Bardeen 
\cite{Bardeen1}. The magnetic and electric fields are given by 
standard definitions, ${\bf B}=\nabla\times{\bf A}$ 
and ${\bf E}=-\nabla\varphi$.

The state of the system is given by the minimum of the total free
energy. Accordingly, variations of ${\mathcal F}$ with respect to the
vector potential $\bf A$, the scalar potential $\varphi$, the GL 
function $\psi$ and the electron density $n$ have to vanish.

Variation ${\delta{\mathcal F}\over \delta {\bar\psi}} =0 $
leads to extended Ginzburg-Landau equation, see \cite{Bardeen1} 
\begin{equation}  \label{eGL}
%                 ================================
{\left(i\hbar\nabla+e^*{\bf A}\right)^2\over 2m^*}\psi
      ={\mu H_c^2\over n} \left(1+
       {t^2\over\sqrt{1-|\psi|^2}}\right)\psi.
\end{equation}
Variation ${\delta{\mathcal F}\over \delta {\bf A}} =0 $
leads to the Maxwell equation,
\begin{eqnarray}  \label{Max1}
%                 ==========================
\nabla\times{\bf B}=-
\mu en{1\over m^*}{\rm Re}\bar\psi(i\hbar\nabla+e^*{\bf A})\psi.
\end{eqnarray}

In principle, density $n$ of the total charge is modified by 
internal electric fields. Accordingly, these equations should 
be treated together with the Poisson equation presented below. 
Since both equations depend on the total density $n$, not on
the deviation from the homogeneous value, it is clear that for
calculating the magnetic properties one can safely neglect the
effect of the small electric field on the density. Within this 
approximation, the set (\ref{eGL}) and (\ref{Max1}) is closed. 

\section{Third GL equation}
%=================================
Variation ${\delta {\mathcal F}\over \delta \varphi}=0$
leads to the Poisson equation,
%\begin{eqnarray}  \label{Max2}
%                 ================================
$-\epsilon\nabla^2\varphi=\rho$.
%\end{eqnarray}
Finally, variation ${\delta{\mathcal F}\over \delta n} =0$
yields the third GL equation,
\begin{eqnarray}  \label{3GL}
%                 ================================
e\varphi
  &=&-\lambda_{TF}^2\nabla^2\varphi
   \nonumber \\
     &+&{1\over 4} {\partial \left(\gamma T_c^2 \right) \over \partial n}
      \left|\psi\right|^2  
     +{1\over2}  T^2{\partial\gamma\over\partial n}
          \sqrt {1-\left|\psi\right|^2}
   \nonumber \\
     &-&{1\over 2} \left(1-{\partial\ln m^* \over \partial\ln n}\right)
        {\left|\left(i\hbar\nabla+e^*{\bf A}\right)\psi\right|^2 \over 2m^*}.
\end{eqnarray}

The first term of (\ref{3GL}) represents the screening on the 
Thomas-Fermi lenght
\cite{JP67}, $\lambda_{TF}^2={\epsilon\over 2{\mathcal N}e^2}$, where 
${\mathcal N}$ is the single-spin density of states. In deriving this 
term we have used the linear response approximation, ${\partial U\over
\partial n}=E_F(n)\approx E_F^0+{\partial E_F\over\partial n}\delta n=
{\partial E_F\over\partial n}{\rho\over e}$, with $E_F^0=0$, and 
employed the Poisson equation. Accordingly, equation (\ref{3GL}) can 
be viewed as the Poisson equation with screening.

The rest of terms in the right hand side of (\ref{3GL}) are source
terms of the third GL equation corresponding to various mechanisms.
The second term is the internal pressure due to the 
dependence of the condensation energy on the total density $n$.
The third term is thermoelectric field of the normal metal reduced 
by factor $\sqrt{1-|\psi|^2}$. The fourth term of is the non-local 
Bernoulli potential \cite{KLB01} but reduced by $|\psi|^2$ in the 
spirit of the so called quasiparticle screening \cite{vS64}. The
correction due to the density dependency of the Cooperon mass, 
$\propto{\partial\ln m^*\over\partial\ln n}$, has been derived by
Rickayzen\cite{Rickayzen} but it is ussually omitted.

\section{Comparison with the BCS results}
%         ===========================
Using the BCS formula for the condensation energy \cite{Tinkham}, 
\begin{equation}
{\cal E}_{\rm con}={1\over 4}\gamma T_c^2=
{1\over 2}\mu H_c^2={1\over 2} {\cal N}\Delta_0^2,
\label{Econdem}
\end{equation}
with $\Delta_0=2\hbar\omega_D \exp(-1/{\cal N}V)$, and the relation 
of the gap to the GL function, $\Delta=\Delta_0\psi$, one recovers 
from the second term of (\ref{3GL}) the BCS expression,
\begin{eqnarray} \label{BCSphi} 
%                 ====================
{1\over 4}{\partial\gamma T_c^2\over\partial n}|\psi|^2
  &\approx& {\Delta^2\over 2}
        {\partial\ln{\mathcal N}\over\partial E_F}
        \ln\left({2\omega_D\hbar\over\Delta_0}\right)
\nonumber\\
  &\approx& {\Delta^2\over 2}{1\over 2 E_F}{1\over {\cal N}V}.
\end{eqnarray}
The first expression has been derived e.g. in \cite{Hong}. 
The second form uses the parabolic band approximation,
${\partial{\mathcal N}\over \partial n}={1\over 2E_F}$.
Together with the estimate, ${\mathcal N}V\sim 1$, it gives the
formula used e.g. in \cite{95Khomskii}.

\section{Conclusion}
%       =============
Starting from the combination of the Gorter-Casimir and 
Ginzburg-Landau free energies further extended by the energy
of the electrostatic field, we have derived a set of three
GL equations. This set includes the extended GL equation 
(\ref{eGL}), the Maxwell equation (\ref{Max1}) for the 
magnetic field and the Poisson equation (\ref{3GL}) for the 
electrostatic potential. In particular limits, the theory
reproduces previous results.

\vspace{5mm}\noindent
This work was supported by GA\v{C}R 202000643, GAAV A1010806 and 
A1010919 grants. The European ESF program VORTEX is also gratefully 
acknowledged. 
%\end{ack}

% References
%================

\end{document}